# Multi-messenger observations of thunderstorm-related bursts of cosmic rays

A. Chilingarian [1], G. Hovsepyan, T. Karapetyan, Y. Khanykyanc, D. Pokhsraryan,

B. Sargsyan, S. Chilingaryan, and S. Soghomonyan

*A. Alikhanyan National Lab (Yerevan Physics Institute), Alikhanyan Brothers 2, Yerevan 0036, Armenia*

*E-mail:* chili@aragats.am

ABSTRACT: We present the facilities of the Aragats Space Environmental Center in Armenia used during multi-year observations of the thunderstorm ground enhancements (TGEs) and corresponding environmental parameters. We analyze the characteristics of the detectors, operated on Aragats, and describe the coordinated detection of TGEs by the network of scintillators, field meters, and weather stations. By using a fast synchronized data acquisition system, we reveal correlations of the multivariate data on time scales from nanosecond to minutes, which allow us to gain insight into the TGE and lightning origin and their interrelations. Also, we demonstrate how different coincidences of multilayered detector operation can select various species of secondary cosmic rays.

---

[1] Corresponding author

**Contents**

### 1. Introduction

The high-energy physics in the atmosphere (HEPA) undergoes a profound transformation in the last decade by performing the correlated measurements of particle fluxes, fast wideband electric field records, and variety of meteorological parameters, including near-surface electric field and geomagnetic field. The synergy of the Cosmic Ray and Atmospheric physics leads to the development of models of the origin of particle bursts registered on the earth's surface, of the vertical profile of the strong electric field in the lower atmosphere, muon stopping effect, interrelations of particle fluxes and lightning flashes, circulation of Radon progeny in the atmosphere, and others. The successes of the multivariate measurements of the last decade put the HEPA to the priority science areas in both the Cosmic Ray and the Atmospheric physics communities. The HEPA research intensified the development of new methods of testing models and theories on atmospheric electricity, particularly in conditions that are related to the most important processes that influence earth environments. Multi-messenger atmospheric science requires, first of all, the development of synchronized networks of identical sensors that are registering the multivariate data and which are stored in databases with open and fast access. The visualization and online correlation analysis of the big data coming from hundreds of measuring channels become a necessary tool in scientific research.

Thunderstorm ground enhancements (TGEs, [1,2]) observed mostly at mountain altitudes by a variety of particle detectors are large impulsive enhancements in the intensity of electrons and gamma rays lasting from tens of seconds to tens of minutes and sometimes enhancing cosmic ray background hundreds of times [3]. If the atmospheric electric field exceeds the critical value, specific to air density (height in the atmosphere), electrons runaway and produce relativistic runaway electron avalanches (RREAs, [4]). The possible configurations of the intracloud electric field that initiated the RREA process are discussed in [5], see Figure 1. Comparison of measured TGEs and simulations with GEANT4 and CORSIKA codes allows outlining the extension and strength of the electric field necessary for starting a runaway process, however, the horizontal extension of the electric field remains still not well understood. Measurements with multiple dosimeters installed at nuclear power plants in a coastal area of the Japanese sea made it possible to follow the source of the gamma ray flux moving with an ambient wind flow [6]. Using the muon stopping effect [7], the size of the particle emitting region was estimated at Nor Amberd research station, located on the slopes of Mt. Aragats at 2000 m height [8]. Estimates from both studies locate particle emitting regions within 1 km. However, in the recent radar-based gamma glow (TGE) study along the coast of the Japanese sea it was observed that all TGEs were accompanied by km long and extended radar echo regions indicating large lower positively charged regions [9]. Thus, the previously considered values of particle emitting region size seem to be highly underestimated. In the present paper, we describe the STAND1 particle detector's network, operated on the Aragats research station aimed to estimate the size of the particle emitting region using a large collection of TGEs registered on Aragats. We are monitoring the secondary cosmic ray fluxes in a 24/7 regime and to avoid ionization losses in building construction, the STAND1 network is located outdoors under deep snow in the Winter season, under a 0.7 mm steel cover. It allows to keep the energy threshold of the upper scintillator ≈1 MeV but poses very stringent conditions on the reliability of detector operation under severe weather conditions.

## 2.     Fast synchronized data acquisition system (FSDAQ)

Most Aragats particle detectors use plastic scintillators with a light-collecting diffuser in the form of a truncated pyramid [10]. At the top of the pyramid, there is a PMT with a large photocathode; at the base, there is usually a 5 cm thick scintillator. The big advantage of such a counter is a good amplitude resolution providing a dynamic range of ≈$10^4$ for the measurements of flux intensity of extensive air shower (EAS) particles hitting the scintillator of a 1 $m^2$ area. The disadvantages include large vertical dimension and mass, and inhomogeneity of light collection at the edges of the scintillator.

Another design of light-collecting scintillator is based on spectrum-shifting re-emitters - fibers glued into the scintillator [11]. Photons of the scintillation flash, are re-emitted to the green part of the spectrum. The attenuation length of light in fiber optic fibers is up to 5 m, thus, the use of this technology allows for the design of the low-cost compact and light particle detectors.

For the STAND1 network, we use 1-$m^2$ sensitive area molded plastic scintillators fabricated by the High Energy Physics Institute, Serpukhov, RF.

The particle data is correlated with the near-surface electrostatic field (NSEF). The Aragats area is continuously monitored by a network of commercially available field mills (Model EFM-100, Boltek Corporation), three of which are placed at the Aragats station, one at the Nor Amberd station at a distance of 12.8 km from Aragats, and one at the Yerevan station, at a distance of 39.1 km from Aragats. The distances between the three field mills at Aragats are 80 m, 270 m, and 290 m. The sensitivity distance of EFM-100 is about 33 km, and the response time of the instrument is 100 ms. The electrostatic field changes are recorded at a sampling interval of 1s and 50 ms. For complete electrical isolation, the field mill is connected to the PC using a fiber optic cable.

The fast synchronized data acquisition (FSDAQ) provides detection of particle fluxes, the near-surface electric field disturbances, and waveforms of radio signals from atmospheric discharges, all harmonized with an accuracy of tens of nanoseconds. In Fig.1 we show two systems of FSDAQ, see details in [12]) operated in 2 experimental halls MAKET and SKL. Both systems employ a 2-channel digital storage oscilloscope (Picoscope 5244B) and National Instrument's NI-myRIO-1900 board [13]. The NI-myRIO board combines the Xilinx Zynq All Programmable SoC with a ready-to-go Linux-based real-time OS (RTOS) is used for synchronization of the data from particle detectors and wideband waveforms registered by electric mills and antennas located on Aragats.

The system in the SKL hall (Fig.1a) includes synchronized measurements of electromagnetic emission produced by atmospheric discharges (fast wideband electric field) and the signal from a particle detector (NaI crystal or 3-cm thick plastic scintillator). The signal from the particle detector is fed to Ch A of the oscilloscope. The fast wideband electric field is registered by a circular flat antenna followed by a passive integrator the output of which is fed to Ch B of the same oscilloscope. The fast wideband electric field measurement system has a bandwidth of ≈50 Hz to 12.5 MHz (RC decay time constant 3 ms). The record length is 1 s, including a pre-trigger time of 200 ms and a post-trigger time of 800 ms. The sampling rate is 25 MS/s (40 ns sampling interval); the amplitude resolution is 8 bits. Starting from 2014, the fast wideband electric field data and particle data are stored on the ASEC servers and are available upon request. FSDAQ located in the SKL hall is triggered by a commercial MFJ-1022 active whip antenna that covers a frequency range from 300 kHz to 200MHz.

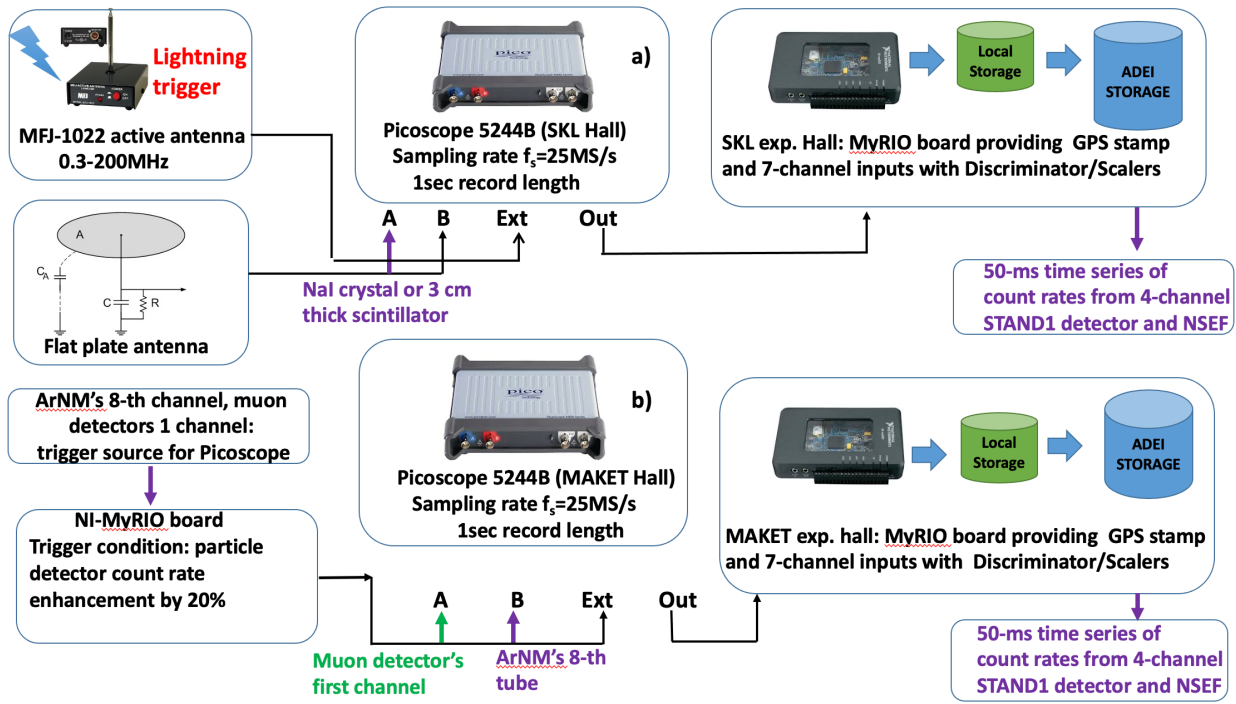

**Figure 1.** Block diagram of two similar systems of the fast synchronized data acquisition (FSDAQ) for the research of particle – lightning relations. Upper system operates in SKL experimental hall, lower – in MAKET hall.

The second FSDAQ system (Fig.1b) located in MAKET experimental hall, is triggered by the abrupt enhancement of the particle flux (particle burst). The signal of the particle detector is fed to the NI MyRio board and, in parallel, to the oscilloscope. When the running mean of particle flux suddenly exceeds the preset background value by 20% the NI MyRio board generates a master pulse for triggering the oscilloscope. The signals from the high-energy muon detector with an energy threshold of 250 MeV and a proportional counter of the Aragats neutron monitor are connected to A and B inputs of Picoscope. For both systems described above, the trigger-out pulse of the oscilloscope is relayed to the National Instruments (NI) MyRIO board which produced the GPS timestamp of the record.

A third MyRio board (without a digital oscilloscope) operates in the GAMMA experimental hall, where the third module of STAND1 detectors is installed.

MyRio boards located in the three experimental halls register the 50-ns time series of the STAND1 modules. At any triggering signal, each of three MyRio boards generates a special output containing the current value of particle detector counts, near-surface electric field value and, a GPS timestamp of the trigger signal. Thus, the fast waveform patterns are synchronized with particle fluxes and with slow (20 Hz) NSEF measurements.

For the location of lightning discharges in Aragats, we use a short-baseline Very High Frequency (VHF) interferometer system [28]. The interferometer (frequency range from 24 to 78 MHz) employs three identical circular flat-plate antennas of 30 cm diameter, which are located in the horizontal plane and form two orthogonal baselines of 13 m in length each. The cross-correlation functions are used to measure the azimuth and elevation angles of the radiation source as a function of time. The interferometer data are digitized at a 156.25 MS/s sampling rate (sample interval of 6.4 ns).

Data on the continuous monitoring of the particle fluxes and environmental parameters are available via advanced data extraction infrastructure (ADEI, http://adei.crd.yerphi.am).

3. **STAND1 particle detector network operated on Aragats station**

In Fig.2 we show the location of the STAND1 network on Aragats station. All three identical units are located nearby three main experimental halls – MAKET, SKL, and GAMMA.

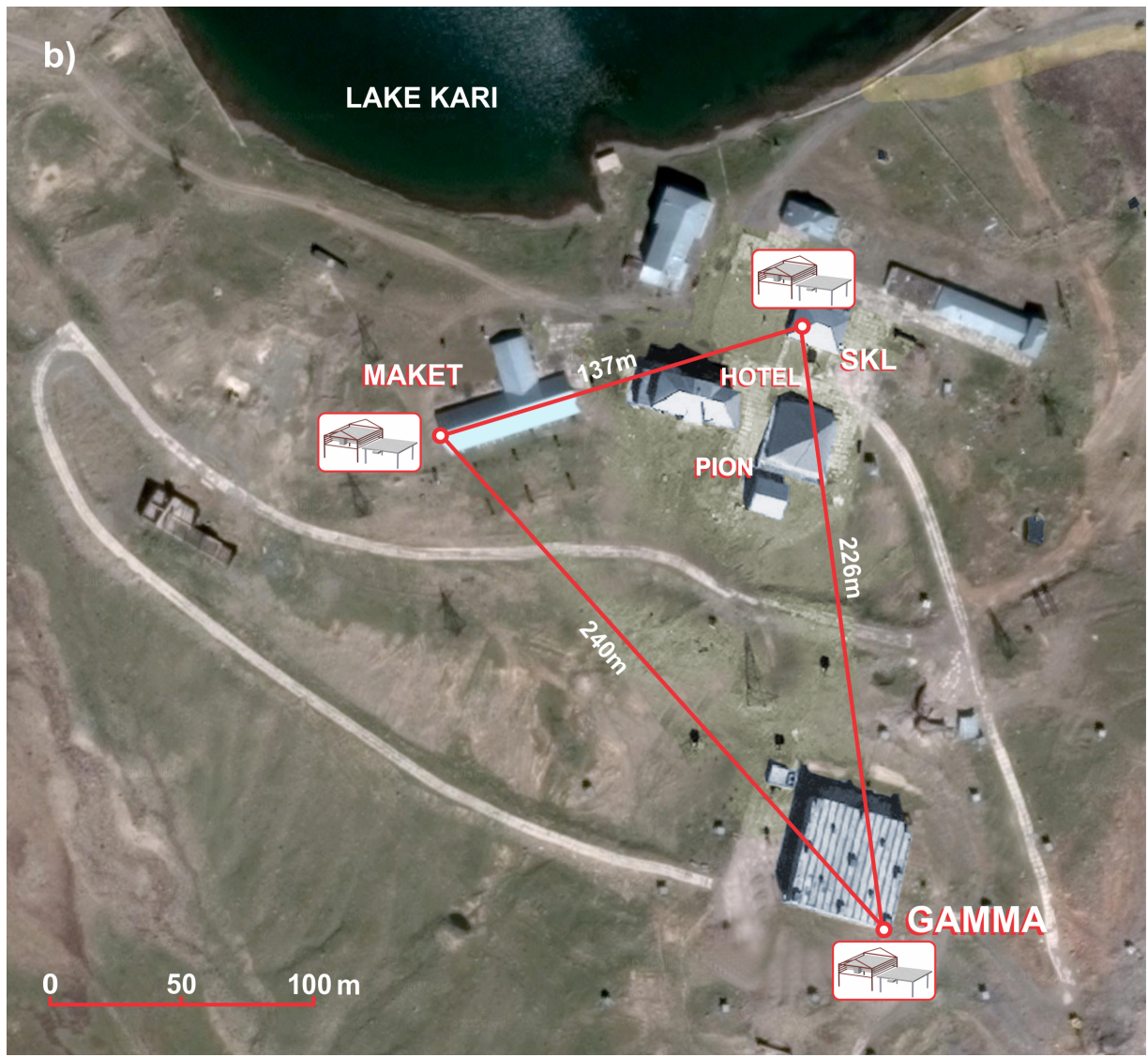

**Figure 2.** Aragats station map with STAND1 network detectors shown.

The "STAND1" detector is comprised of three layers of 1-cm-thick, 1-m² sensitive area scintillators stacked vertically and one 3-cm thick plastic scintillator of the same type stands apart; see Fig. 3. The light from the scintillator through optical spectrum-shifter fibers is reradiated to the long-wavelength region and passed to the photomultiplier FEU-115M. The maximum luminescence is emitted at the 420-nm wavelength, with a luminescence time of about 2.3 ns. The STAND1 detector is tuned by changing the high voltage applied to the photomultiplier (PM) and setting the thresholds for the discriminator shaper. The discrimination level is chosen to guarantee both high efficiency of signal detection and maximal suppression of photomultiplier noise.

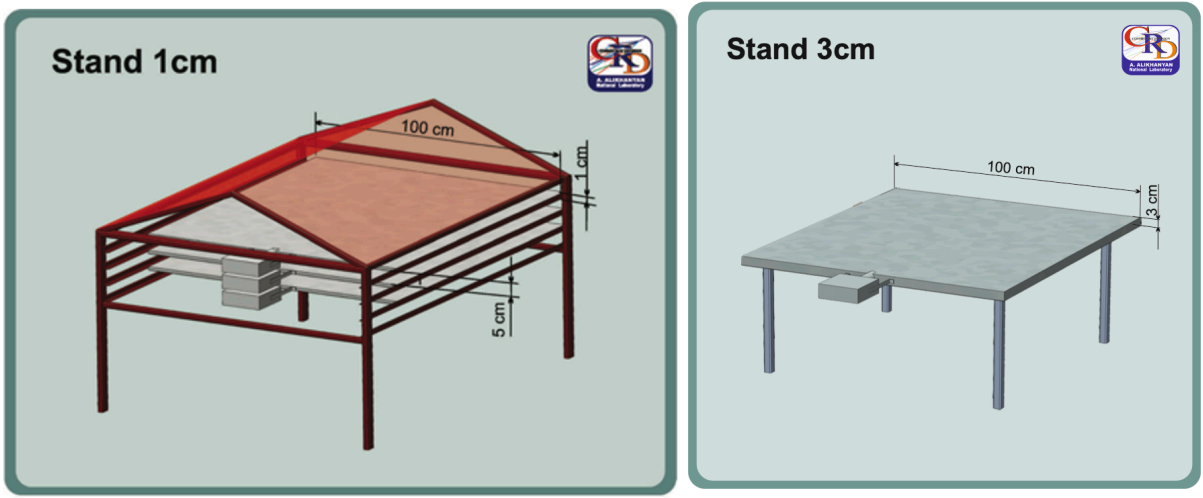

**Figure 3.** STAND1 detector consisting of three layers of 1-cm- thick scintillators (left) and stand-alone 3-cm thick plastic scintillator of the same type (right).

In Fig. 4 we show the typic shape of the pulse from the 1-cm thick scintillator of the STAND1 detector. As is expected, the pulse is quite short, the full width on half maximum (FWHM) is less than 30 ns, and the maximum amplitude is reached in a few nanoseconds. It makes the detector very suitable for correlation studies with fast phenomena like lightning flashes.

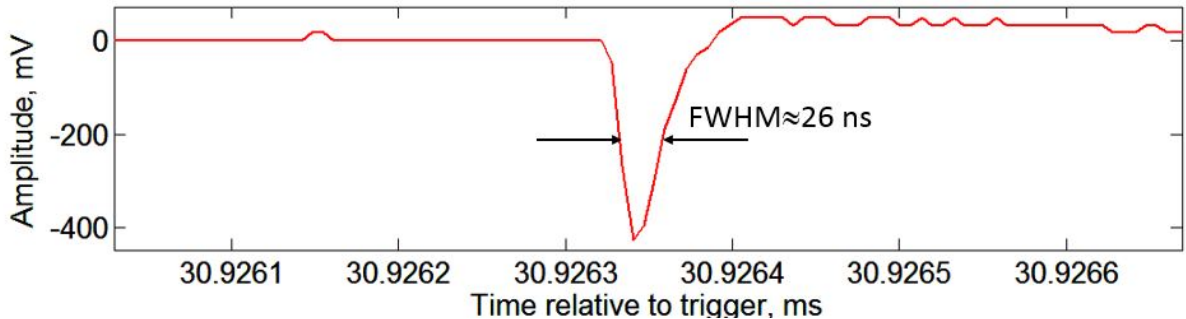

**Figure 4.** Typical shape of the pulse from single particle registered by 1-cm thick scintillator.

In Fig. 5 we show the response of the 1-cm thick scintillator to 10-MeV gamma rays and electrons obtained by GEAN4 simulations.

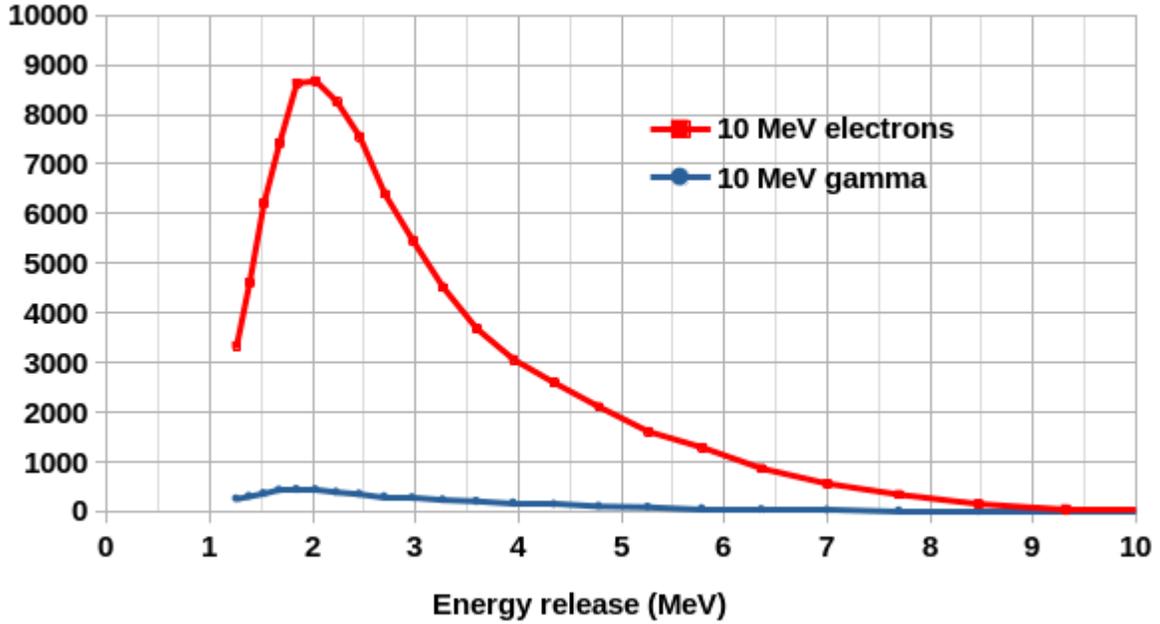

**Figure 5** . The response function of the 1-cm thick scintillator of STAND1 detector to 10-MeV electrons (100,000 simulation trials) and 10-MeV gamma rays (500,000 simulation trials, normalized to 100,000).

As we can see in Fig. 5 the efficiency of electron registration is much larger than the efficiency of gamma ray registration. Values obtained from simulations are ≈95% for electrons and ≈3% for gamma rays. The modes of distribution are peaked at ≈2 MeV (electrons) and at ≈1.76 MeV (gamma rays). Thus, the 1-cm thick scintillators are perfectly suited for the measurement of a number of avalanche particles in the detector, by the analysis of pulse amplitudes. The 30-fold suppression of the gamma rays is also a very useful feature in the TGE research. Due to ionization losses in the air, the majority of TGE particles are gamma rays; the coincidences of STAND's 1-cm scintillators, along with the ASNT spectrometer [29], allow us to separate usually very weak electron flux from the gamma ray flux.

The operation of the STAND1 network was rather stable in Spring 2018. The mean values of 50-ms count rates from 2 May to 2 June were 23.7±1.65 (GAMMA) and 24±0.9 (MAKET), and 20±3 (SKL) relative errors correspondingly 7, 2.7%, and 15% that is not large for such a short time sampling.

4. **Synchronized registration of particle fluxes and radio emission from the atmospheric discharges**

Synchronization of the STAND1 modules in different experimental halls by the FSDAQ electronics was checked with correlation analysis between all 3 remote modules. In Fig.6d we show 1-s time series of the upper scintillator of the STAND1 network from 18:28 to 18:34 (a total of 6 minutes). We can see a more-or-less smooth enhancement of the count rate by 67% (≈10σ) above the fair-weather value: the count rate increases from 460 to 767 counts per second (for the unit located on the roof of the GAMMA calorimeter, black curve). A large TGE was terminated by a lightning flash on a maximum phase of its development at 18:33.22:275 UT. Count rate time series demonstrate coherent enhancement and termination, and in the scatter plots in the upper panel of Fig. 6(a-c) we see strong correlations of count rates measured by three independent scintillators. Thus, the electron accelerator above the research station is operating very stable during 5 minutes of TGE, sending a large flux of electrons and gamma rays in the direction to the earth's surface.

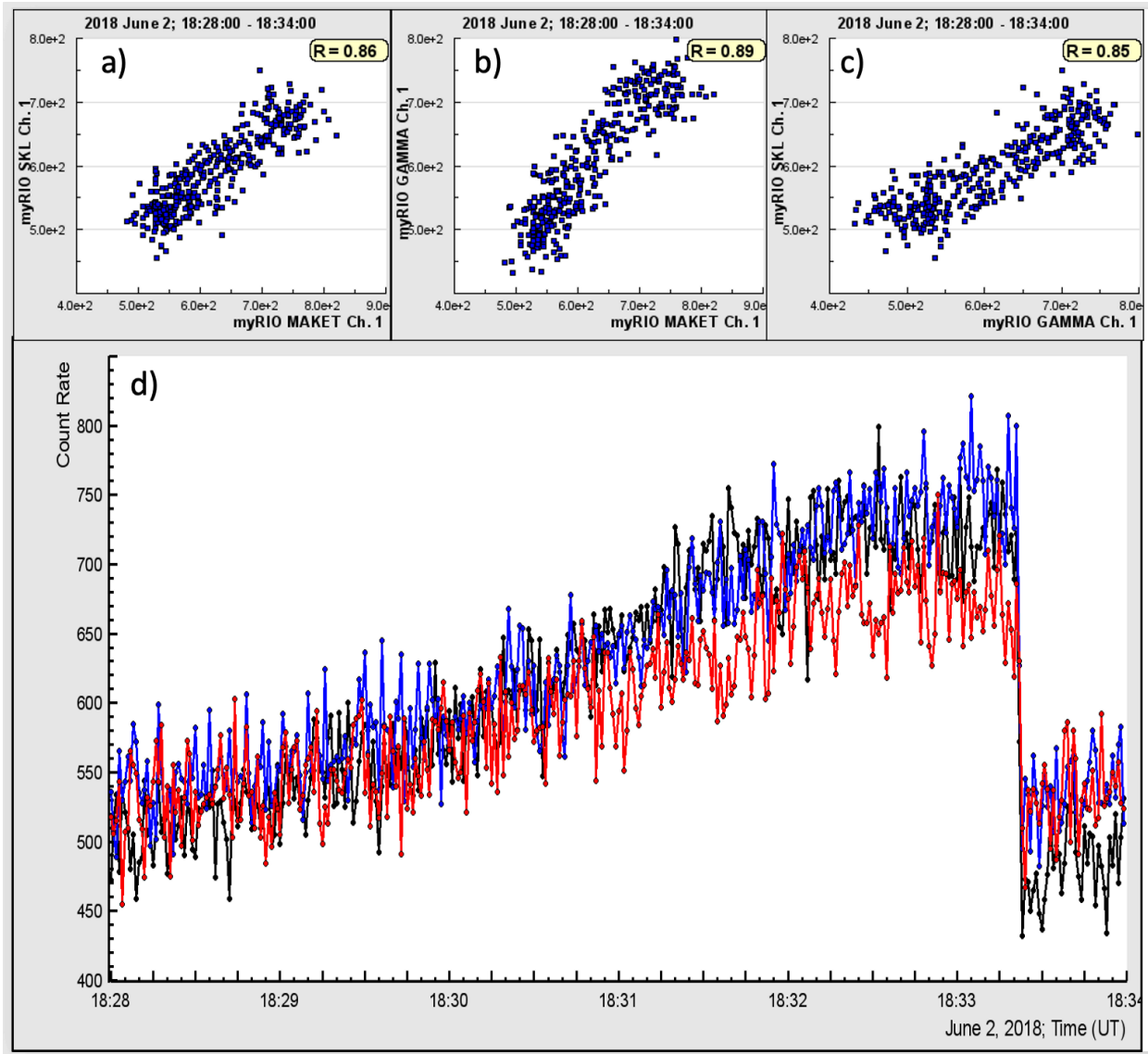

**Figure 6.** 1-s time series registered by the STAND1 nework: GAMMA (black), MAKET (blue), SKL (red), for unit locations and distances see Fig.3. In the upper panel – all 3 scatter plots of possible combination of network scintillator with calculated correlation coefficients.

However, 1-s time synchronization shown in Fig. 6 is not enough for the joint analysis of particle fluxes and lightning occurrences usually terminated the particle flux, see collection of 165 TGEs abruptly terminated by a flash in the Mendeley data set [14]. In Fig. 7 and Table 1 we show further analysis of TGE, now on 50-ms time scale, a total of 10 s. We can see that the abrupt termination of the TGE seen on 1-s time scale in Fig. 7 is coherent also on the time scale of 50-ms. However, the shapes and amplitudes of the disturbances of the NSEF are different, as well as the count rates of the 3 scintillators, see Table 1.

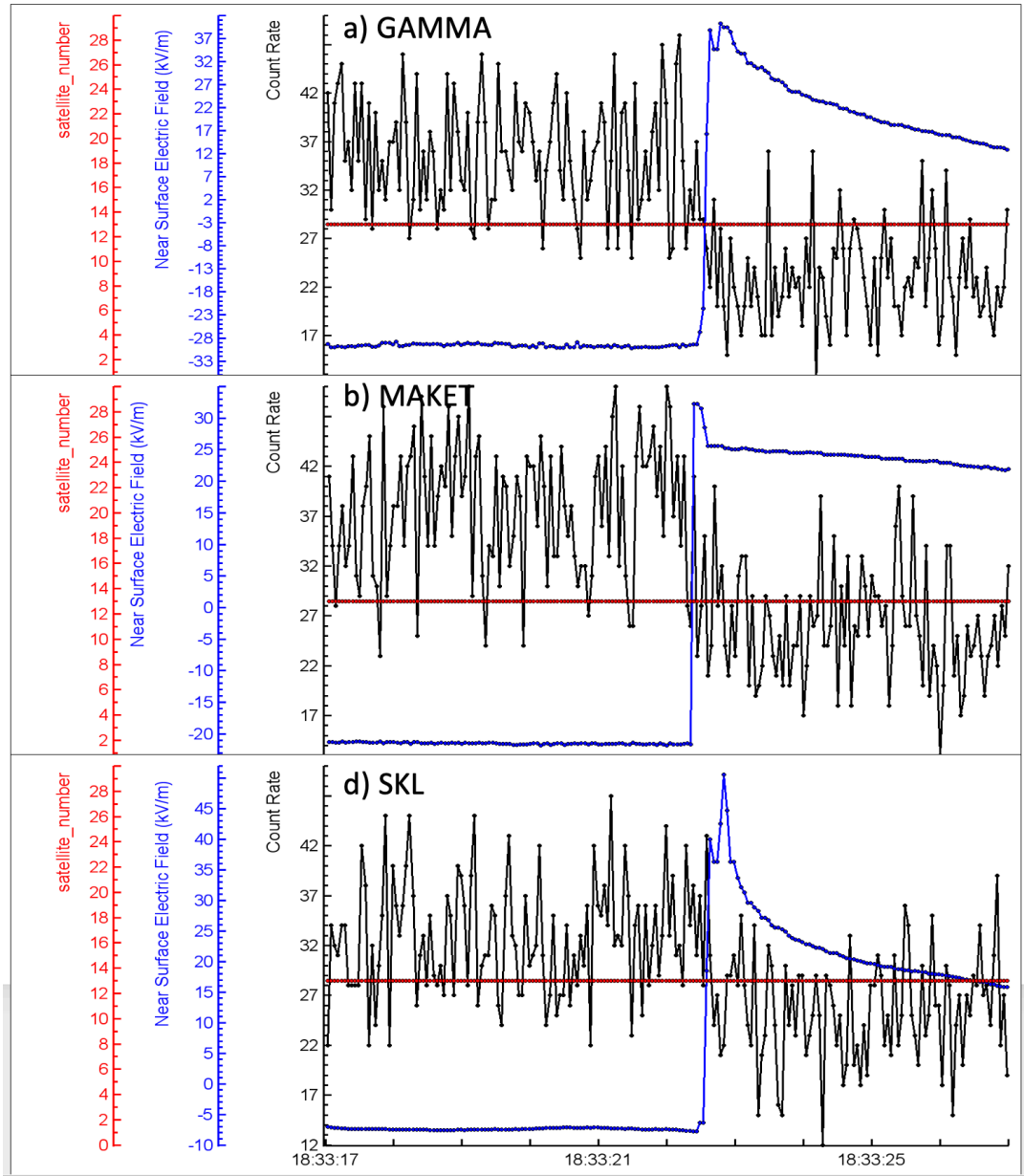

**Figure 7.** 50-ms time series of STAND1 network scintillators (black), disturbances of the NSEF, registered by the EFM-100 sensors (blue), and number of satellites used for the determination of GPS time stamps (red, 13 satellites operational for all sites)..

In the second and third columns, we show the mean values of 50-s count rates before and after lightning flash, in the fourth column the surge of particle count rate, in the fifth and sixth – the NSEF measurements before the lightning flash, and at its local maximum. In the last column - the duration of the NSF surge from start to local maximum.

The differences in the local measurements of the NSEF are expected due to different locations on the mountain terrain. As was mentioned in [15] and in the references therein, "the mountain geometry can strengthen the electric field by a factor on the order of 2". The location of NSEF sensors differs by the height of the mast on which they are attached to be out of the deep snow during winter months and the mountain environments also are different in different locations. Sure, the scintillators are also not fully identical, and the high-voltages and discrimination thresholds can be changed during multi-year operation. Also, obviously, we are not dealing with man-made accelerators, with fixed bunches of particles, The atmospheric electron accelerator sends to the earth multiple RREA avalanches, that differ in acceleration distances, electron, and gamma ray numbers, and maximum energies. A huge number of such avalanches constitute the enhanced particle fluxes that reach the earth's surface, and these enhanced fluxes can be stable on a second or even on a minute time scale. More detailed consideration of the particle fluxes and the disturbances of NSEF reveals complicated shapes and inhomogeneities seen on ms time scales.

**Table 1. Parameters of particle fluxes and disturbances of NSEF registered during 10 s of maximum flux and termination of TGE, shown in Fig. 7.**

| 18:33:17-18:33:27 | Mean before flash | Mean after flash | Surge of count rate | NSEF before flash (kV/m) | NESF after flash (kV/m) | NSEF surge | Duration of surge (ms) |
|---|---|---|---|---|---|---|---|
| GAMMA | 29±8 | 20±11 | 9 (44%) | -29 | 38 | 67 | 200 |
| MAKET | 38±7 | 26±6 | 12(32%) | -21 | 32 | 53 | 50 |
| SKL | 33±6 | 25±5 | 8 (24%) | -7 | 50 | 57 | 100 |

Further zooming of synchronized detection of the pulses from particle detectors and atmospheric discharges is possible with high-speed oscilloscopes, using GPS timestamps produced by the NI MyRIO (FSDAQ system). For many years on Aragats, we continue experiments to find particle bursts generated during the lightning flash. Despite a few observations reported in [16-20], we didn't register any particle bursts coinciding with lightning flashes. Numerous lightning flashes, registered on Aragats, visa verse, terminate TGEs, not originate them [14]. Our observations, as well a number of largest cosmic ray experiments relate registered short particle bursts to well-known physical phenomena, namely extensive air showers [21-25]. To resolve this contradiction, and detect the possible generation of high-energy particles (with tens of MeV energies) in the lightning bolt, we measure simultaneously signals from a variety of particle detectors (NaI crystals, proportional chambers of neutron monitor, scintillators of different thicknesses, and area) simultaneously with electromagnetic radiation from the atmospheric discharges (see details in [26]). After multiyear observations, the only coincidences we detect were electromagnetic interferences (EMI). A lightning flash generates tens of kA currents in the atmosphere and, therefore, powerful electromagnetic radiation. If flash is nearby particle detectors it is rather difficult to screen it from this huge source of noise. Thus, some particle detectors generate pulses that can mimic the particle signals. However, if we look at the patterns of both EMI registered by antennas and particle output signals, we easily distinguish EMI from genuine particle bursts. In Fig.8 we show a 70-μs fragment of the synchronized registration of one of the NaI detector's signals and the signals of the flat plate antenna by the same oscilloscope. The accuracy of synchronization is 40 ns (sampling interval of digitized signals).

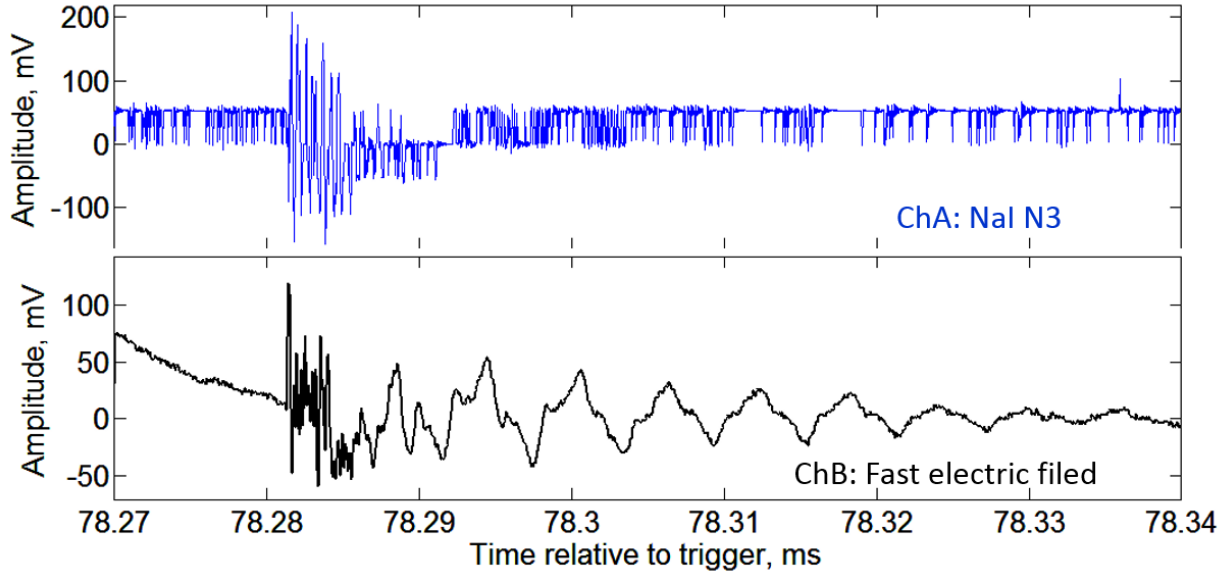

**Figure 8.** 40-ns time series of NaI spectrometer N 3 of 7 crystal network operated in the SKL experimental hall (blue, channel A of picoscope), and the signals from the flat-plate antena registering the electromagnetic radiation from atmospheric discharges (black, channel B of picoscope)..

We can see that NaI pulses (upper panel) ideally correlate with the fast electric field pulses (lower panel) and are bi-polar, whereas the genuine pulses from the particles registered by the large NaI crystal are unipolar and negative, see Fig 9. Thus, the "particle burst" registered by NaI, in fact, is an EMI signal induced in the detector circuit.

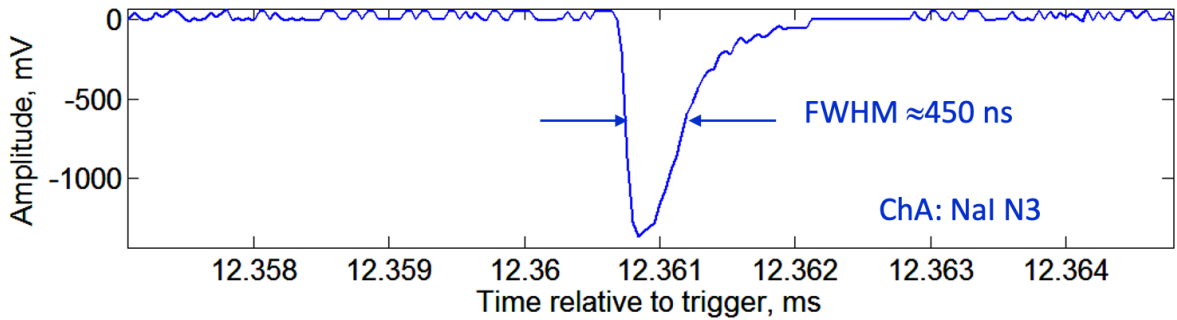

**Figure 9.** Typical shape of the signal generated by a particle in the large NaI crystall (12 x 12 x 24 cm).

## 5. Simulation of the STAND1 detector response function. Coincidences of detector layers.

The STAND1 detector located nearby the MAKET experimental is connected (in parallel to the NI MyRIO board) also to alternative electronics which makes it possible to register not only the count rates of detector layers but as well, various combinations of their possible coincidences. In this way, we select different species of the ambient cosmic ray flux for further analysis. In Table 2 and Fig 10 we show the results of detector response calculation with fluxes obtained from the EXPACS WEB calculator [27]. As we can see the "001" and "010" coincidences are enriched by neutral particles, and the "111" coincidence selects charged particles. In the last column of Table 2, we post the one-minute count rate of every coincidence assuming the energy threshold of all 3 scintillators to be 1.2 MeV. The corresponding 1-minute count rates of STAND1 layers are 31600, 29500, and 27600.

**Table 2. The share of each of the species of cosmic ray bagkround flux "selected" by different coincidences of the STAND1 scintillators**

| Coincidence | Neutron % | Proton % | mu+ % | mu- % | Electron % | Positron % | Gamma % | count rate |
|---|---|---|---|---|---|---|---|---|
| **001** | 20.46 | 0.56 | 4.02 | 3.57 | 1.90 | 1.96 | 67.53 | 4429.0 |
| **111** | 0.43 | 7.07 | 36.00 | 31.65 | 11.12 | 9.91 | 3.90 | 20149.0 |
| **110** | 3.10 | 4.77 | 17.51 | 15.16 | 25.59 | 17.49 | 15.94 | 3877.0 |
| **100** | 17.10 | 4.20 | 5.85 | 5.43 | 23.65 | 1376 | 30.01 | 7418.0 |
| **011** | 5.66 | 1.27 | 16.31 | 14.65 | 5.54 | 5.46 | 51.10 | 2581.0 |
| **101** | 4.10 | 1.12 | 36.56 | 31.70 | 12.77 | 11.06 | 2.70 | 1245.0 |
| **010** | 31.24 | 0.17 | 0.99 | 0.96 | 2.59 | 1.85 | 62.17 | 3431.0 |

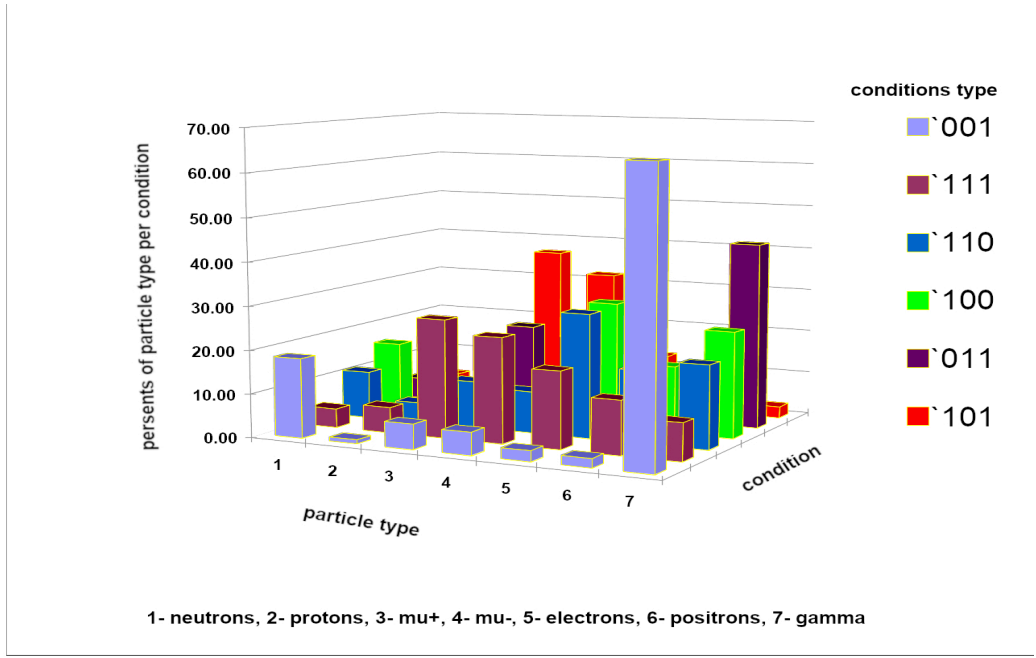

**Figure 10.** Graphical presentation of data posted in Table 2.

From the ADEI data analysis platform [30] we can readily obtain both online and archived data in graphical and numerical forms. In Table 3 we post the background and TGE peak minute values of several STAND1 coincidences. We choose the coincidences related to the lower layer for the analysis of the highest energies of TGE particles. First of all, we check that the sum of the coincidences, which have been registered in the lower layer (namely 101, 011, 111, and 001) equals the count registered by the STAND1 lower layer. Then, we can examine the coincidences by comparing them with the data from Table 2. For instance, we can see that 001 coincidence selects ≈95% TGE gamma rays and only with 5% probability electrons and positrons (other species of cosmic rays will not enhance during TGE). Thus, the enhancement of 001 coincidence can be related to the TGE gamma rays. From Figure 11 ( from [31]) we can see that the energies of gamma rays selected by the 001 coincidence are larger than 10 MeV. Thus, we get an estimate of large energy gamma ray content in the TGE.

**Table 3. The maximum and backround values of the minute count rates of STAND1 coincidences for the TGE that occurred at 18:33 on June 2, 2018.**

| Count rate | STAND1 Lower | STAND1 101 | STAND1 011 | STAND1 111 | STAND1 001 |
|---|---|---|---|---|---|
| **TGE +background** | 20500 | 286 | 2469 | 12572 | 5173 |
| **Background** | 17199 | 233 | 2116 | 10797 | 4053 |
| **TGE** | 3301 | 53 | 353 | 1785 | 1170 |

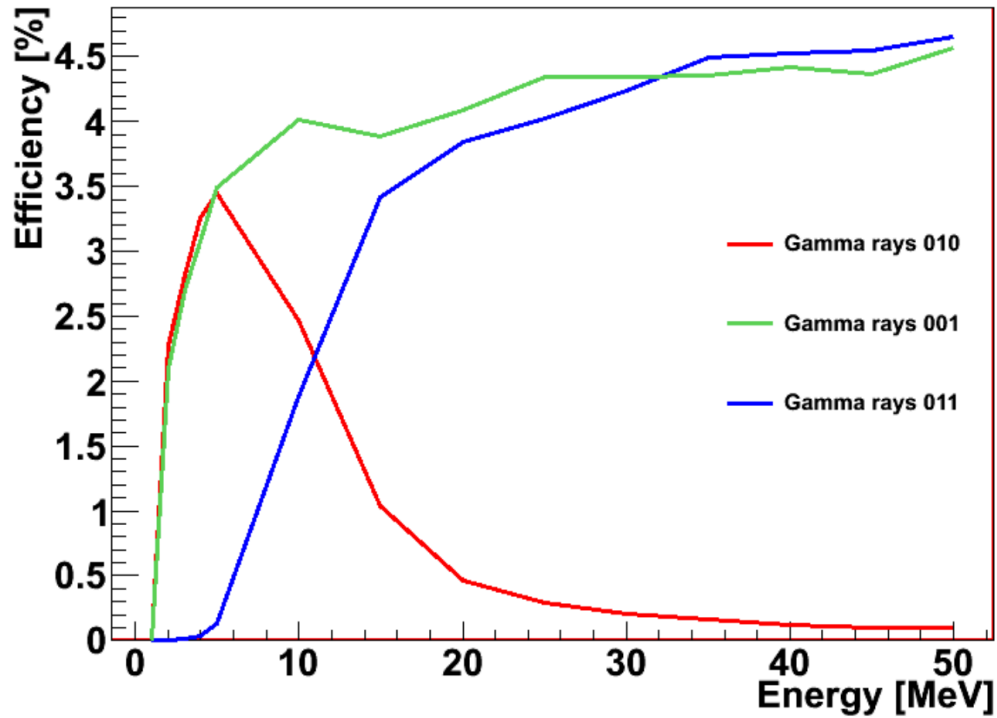

**Figure 11.** The energy dependence of different STAND1 coincidences to register gamma rays.

Comparing STAND1 coincidences with registration efficiencies and shares of each species we can estimate the content of other particles in TGE. If the TGE is large, as we can see in Fig. 12 (100 coincidence's enhancement reaches 55%), we recover energy spectra of charged and neutral components of TGE with the ASNT Spectrometer [29]. Afterward, we can obtain by simulation count rates of all coincidences by applying recovered electron and gamma ray fluxes to the STAND1 detector assembly.

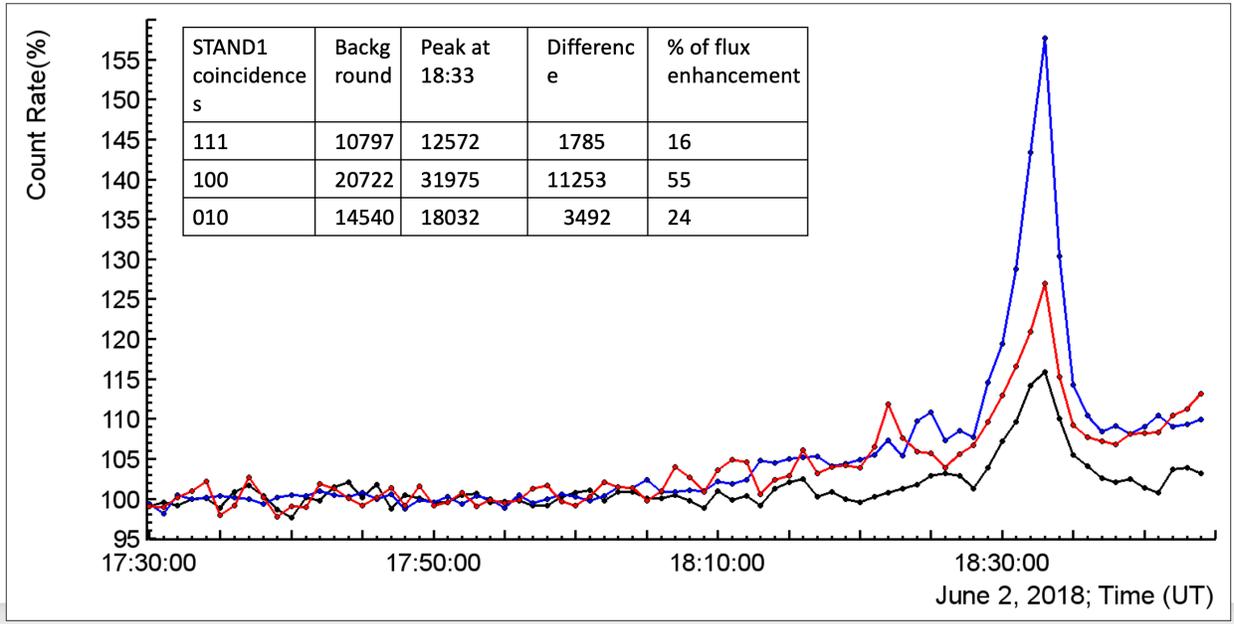

**Figure 12.** 1-minute time series of count rates of STAND1 coincidences.

In Fig. 13 we demonstrate another possibility of the analysis of the STAND1 detector's coincidences. By dividing counts of 101 coincidence by 111 coincidence counts at fair weather and during TGE we estimate the efficiency of the middle scintillator to be 97-98%.

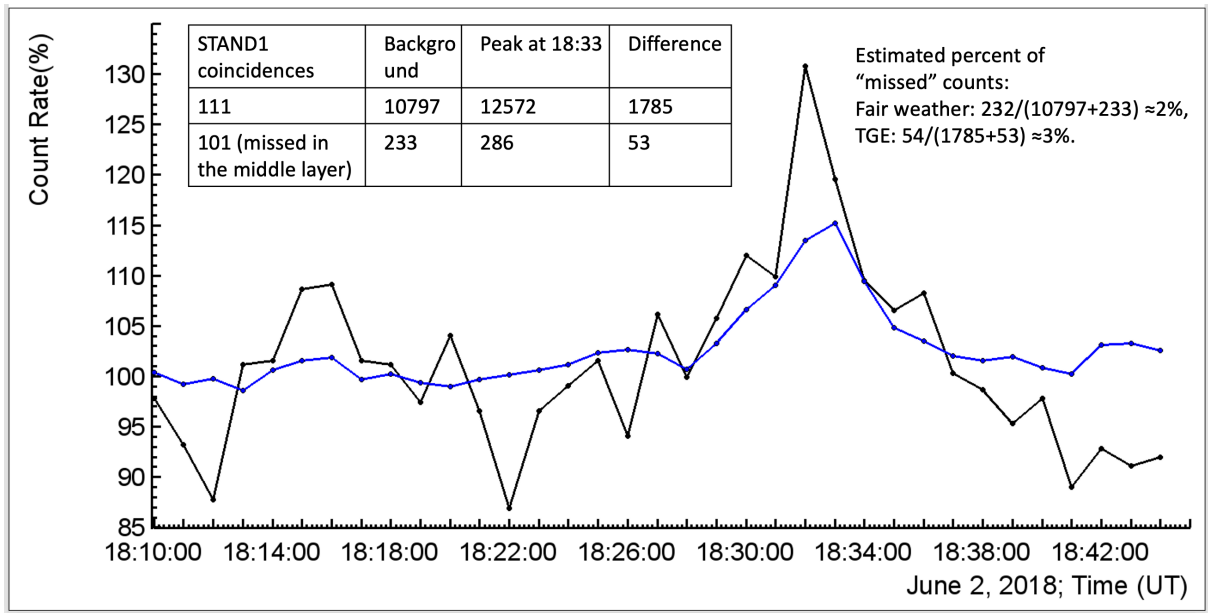

**Figure 13.** Comparison of efficiences of the middle scintillator of STAND1 at fair weather and during TGE. Black curve – 101 coincidence, blue curve 111 coincidence.

In Fig. 14, we show one more possibility of STAND1 data analysis. As we see from the Figure, the enhancement of the gamma ray flux (selected by 001 coincidence) started at 17:45 long before the TGE maximum at 18:33. The disturbances of the NSEF just started and cannot initiate TGE, however, the $^{226}$Radon progeny, which are attached to charged aerosols are lifted by the NSEF and cause enhancement of the count rate of the lower STAND1 detector's scintillator [32].

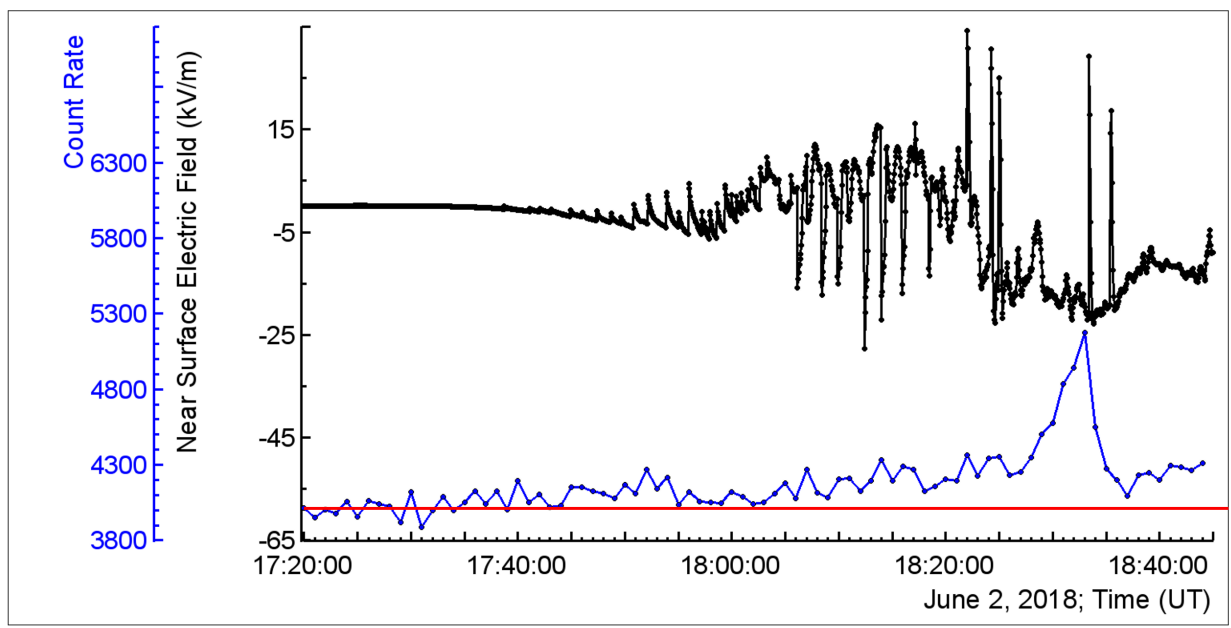

**Figure 14.** Time series of disturbances of NSEF (black curve) and count rate of 001 STAND1 coincidence. By the red line we show that the enhancement of count rate started at 17:45 long before the TGE start.

**Conclusions**

The progress in the High Energy Physics in the Atmosphere (HEPA) research has been pushed by establishing the networks of the same type of particle detectors at different sites at Aragats Space Environmental Center and in the countries of Eastern Europe and Germany and Japan. The joint operation of these networks has shown that TGEs is a universal process sharing the same characteristics measured at different observation sites. The correlation analysis of the enhanced particle fluxes registered by these networks can be used for the research of the radiation-emitting region in the thundercloud. The comparisons of the delayed correlations between the signals from the network of scintillation detectors and local disturbances of the near-surface electric field measured by electric field mills open the possibility of better understanding the vertical and horizontal profiles of the atmospheric electric field. Analysis of the 1-minute time series of STAND1 coincidences allows us to identify the TGE event and estimate its significance and the content of electrons. Registered coincidences of STAND1 layers give additional possibilities to separate different species of the cosmic ray flux, for researching the muon stopping effect, neutron production during thunderstorms, and others (see Tables 2 and 3, and Figs. 10-14). The correlations of STAND1 modules during large TGE on 2 June (Fig.6) demonstrate that at least over an area of ≈ $10^6$ m2 fluxes of TGE particles are identical and, therefore vertical profile of the atmospheric electric field can sustain fields strengths of 2.0 – 2.2 kV/cm for minutes. On milliseconds time scales we can observe inhomogeneities of the NSEF reflecting the approaching storm and peculiarities of mountain terrain. Particle bursts observed on Aragats can be explained by the conventional EAS physics (Figs. 8 and 9). Thus, EAS physics and HEPA are synergistically connected and need to exchange results for the explanation of particle bursts and for revealing the influence of atmospheric electric fields on the EAS shape and size.

The correlations of NSEF with particle fluxes possibly will help to get insight into enigmatic TGE-lightning relations (Fig. 7, Table 1).

**Acknowledgments**

We thank the staff of the Aragats Space Environmental Center for the uninterruptible operation of experimental facilities on Aragats under severe weather conditions. The data for this study is available in numerical and graphical formats by the multivariate visualization software platform ADEI on the WEB page of the Cosmic Ray Division (CRD) of the Yerevan Physics Institute, http://adei.crd.yerphi.am/adei and from Mendeley datasets [14,26]. The authors acknowledge the support of the Science Committee of the Republic of Armenia (research project No 21AG- 1C012), in the modernization of the technical infrastructure of high-altitude stations.

**Declaration of Competing Interest**

The authors declare no conflict of interest.

**Data Availability Statement**

The data for this study are available in numerical and graphical formats on the WEB page of the Cosmic Ray Division (CRD) of the Yerevan Physics Institute, http://adei.crd.yerphi.am/adei and from Mendeley datasets [14,26].

**References**

1. Chilingarian A., Daryan A., Arakelyan K. et al (2010) Ground-based observations of thunderstorm-correlated fluxes of high-energy electrons, gamma rays, and neutrons. Phys Rev D 82:043009
2. Chilingarian A., Hovsepyan G., and Hovhannisyan A., Particle bursts from thunderclouds: Natural particle accelerators above our heads, Physical review D 83, 062001 (2011).
3. Chum, R. Langer, J. Baše, M. Kollárik, I. Strhárský, G. Diendorfer, J. Rusz, Significant enhancements of secondary cosmic rays and electric field at high mountain peak during thunderstorms, Earth Planets Space 72 (2020) 28.
4. Gurevich A.V., Milikh G.M., Roussel-Dupre R. Runaway electron mechanism of air breakdown and preconditioning during a thunderstorm. Physics Letters A. 1992; 165(5–6): 463–468. DOI:10.1016/0375-9601(92)90348-P.

5. A.Chilingarian, G. Hovsepyan, and M. Zazyan, Measurement of TGE particle energy spectra: An insight in the cloud charge structure, Europhysics letters (2021), 134, 6901, https://doi.org/10.1209/0295- 5075/ac0dfa
6. Torii, T., Sugita, T., Kamogawa, M., Watanabe, Y., & Kusunoki, K. Migrating source of energetic radiation generated by thunderstorm activity. Geophysical Research Letters , 38 (24).
7. A.Chilingarian, G. Hovsepyan, G.Karapetyan, and M.Zazyan, Stopping muon effect and estimation of intracloud electric field, Astroparticle Physics 124 (2021) 102505.
8. A Chilingarian, N Bostanjyan, T Karapetyan, On the possibility of location of radiation-emitting region in thundercloud, IOP Publishing Journal of Physics: Conference Series **409** (2013) 012217 doi:10.1088/1742-6596/409/1/012217
9. Wada, Y., Enoto, T., Kubo, M., Nakazawa, K., Shinoda, T., Yonetoku, D., et al. (2021). Meteorological aspects of gamma-ray glows in winter thunderstorms. *Geophysical Research Letters*, *48*, e2020GL091910. https://doi. org/10.1029/2020GL091910
10. A.Chilingarian, G. Gharagyozyan, G. Hovsepyan, S. Ghazaryan, L. Melkumyan, A. Vardanyan, Light and heavy cosmic-ray mass group energy spectra as mea- sured by the MAKET-ANI detector, Astrophys. J. 603, L29 (2004).
11. G.I. Britvich, S.K. Chernichenko, A.P. Chubenko, et.al. Nucl. Instr. and Meth. A 564 (2006) 225-234.
12. D. Pokhsraryan, Fast Data Acquisition system based on NI-myRIO board with GPS time stamping capabilities for atmospheric electricity research, Proceedings of TEPA symposium, Nor Amberd, 2015, p.23, "Tigran Mets", Yerevan, 2016.
13. https://zone.ni.com/reference/en-XX/help/373197L-01/myriodriverhelp/ni_myrio-1900/
14. Soghomonyan, Suren; Chilingarian, Ashot, Khanikyants, Yeghia (2021), "Dataset for Thunderstorm Ground Enhancements terminated by lightning discharges", Mendeley Data, V1, doi:10.17632/p25bb7jrfp.1 https://data.mendeley.com/datasets/p25bb7jrfp/1
15. Hager, W. W., and W. Feng (2013), Charge rearrangement deduced from nearby electric field measurements of an intracloud flash with K-changes, J. Geophys. Res. Atmos., 118, 10,313–10,331, doi:10.1002/jgrd.50782.
16. Mallick, S., V. A. Rakov, and J. R. Dwyer (2012), A study of X-ray emissions from thunderstorms with emphasis on subsequent strokes in natural lightning, J. Geophys. Res., 117, D16107, doi:10.1029/2012JD017555.
17. M.D. Tran, V.A. Rakov, S. Mallick, et al., A terrestrial gamma-ray flash recorded at the Lightning Observatory in Gainesville, Florida, Journal of Atmospheric and Solar-Terrestrial Physics 136 (2015) 86–93.
18. J. R. Dwyer, H. K. Rassoul, M. Al-Dayeh, et al., A ground level gamma-ray burst observed in association with rocket-triggered lightning, GEOPHYSICAL RESEARCH LETTERS, VOL. 31, L05119, doi:10.1029/2003GL018771, 2004.
19. J.W. Belz, P.R. Krehbiel, J. Remington et al., Observations of the Origin of Downward Terrestrial Gamma-Ray Flashes, JGR, Atmospheres, 125, e2019JD031940 (2020).
20. Wada, Y., Morimoto, T., Nakamura, et al. (2022). Characteristics of low-frequency pulses associated with downward terrestrial gamma-ray flashes. *Geophysical Research Letters*, *49*, e2021GL097348. https://doi. org/10.1029/2021GL097348

21. A.U. Abeysekara, J.A.Aguilar, S. Aguilar, S., et al., On the sensitivity of the HAWC observatory to gamma-ray bursts. Astroparticle Physics, 35, 641, (2012). https://doi.org/10.1016/j.astropartphys.2012.02.001
22. Z. T.Izhbulyakova, A. G. Bogdanov, F. A. Bogdanov, et al., Investigation of the EAS neutron component with the URAN array: first simulation and experimental results, Journal of Physics: Conference Series 1690 (2020) 012071 doi:10.1088/1742-6596/1690/1/012071
23. A.P.Chubenko, A.L.Shepetov, V.P.Antonova, et al., New complex EAS installation of the Tien Shan mountain cosmic ray station, NIM, 832, 158 (2016) https://doi.org/10.1016/j.nima.2016.06.068
24. B. Bartoli, P. Bernardini, X.J. Bi et.al., Detection of thermal neutrons with the PRISMA-YBJ array in Extensive Air Showers selected by the ARGO-YBJ experiment, Collaborations, *Astropart.Phys. 81 (2016) 49-60*
25. A.Chilingarian, S. Soghomonyan, Y. Khanikyanc, D. Pokhsraryan, On the origin of particle fluxes from thunderclouds, Astroparticle Physics 105, 54 (2019).
26. Soghomonyan, Suren; Chilingarian, Ashot; Pokhsraryan, David (2021), "Extensive Air Shower (EAS) registration by the measurements of the multiplicity of neutron monitor signal", Mendeley Data, V1, doi: 10.17632/43ndcktj3z.1 https://data.mendeley.com/datasets/43ndcktj3z/1

27. T. Sato, Analytical model for estimating the zenith angle dependence of terrestrial cosmic ray fluxes, PLOS ONE 11 (2016) e0160390.

28. A. Chilingarian, M. Dolgonosov, A. Kiselyov, Y. Khanikyants and S. Soghomonyan, Lightning observations using broadband VHF interferometer and electric field measurements, Journal of Instrumentation 15, P07002, (2020)
29. A. Chilingarian, G. Hovsepyan, T.Karapetyan, et al., Measurements of energy spectra of relativistic electrons and gamma-rays avalanches developed in the thunderous atmosphere with Aragats Solar Neutron Telescope, Journal of Instrumentation, 17, P03002 (2022).
30. Chilingaryan, S. *et al*. The Aragats data acquisition system for highly distributed particle detecting networks. *Journal of Physics: Conference Series* **119**, 082001 (2008).
31. Mailyan B., the efficiencies of the Aragats space environmental center (ASEC) particle detectors used in thunderstorm ground enhancement (TGE) research, ADEI WiKi, http://adei.crd.yerphi.am/setups/asec/pictures/Efficiencies.pdf
32. Chilingarian, A., Hovsepyan, G., & Sargsyan, B. (2020). Circulation of Radon progeny in the terrestrial atmosphere during thunderstorms. *Geophysical Research Letters*, *47*, e2020GL091155. https://doi. org/10.1029/2020GL091155.